\documentclass[conference]{IEEEtran}
\usepackage{multirow} 
\IEEEoverridecommandlockouts
\usepackage{cite}
\usepackage{hyperref}
\usepackage{amsmath,amssymb,amsfonts}

\usepackage{graphicx}
\usepackage{textcomp}
\usepackage{xcolor}
\def\BibTeX{{\rm B\kern-.05em{\sc i\kern-.025em b}\kern-.08em
    T\kern-.1667em\lower.7ex\hbox{E}\kern-.125emX}}
\usepackage{algorithm}
\usepackage{algpseudocode}
\begin{document}

\title{M-SCAN: A Multistage Framework for Lumbar Spinal Canal Stenosis Grading Using Multi-View Cross Attention\\
}

\author{\IEEEauthorblockN{Arnesh Batra*}
\IEEEauthorblockA{\textit{IIIT-Delhi} \\
Delhi, India \\
arnesh23129@iiitd.ac.in}
\and
\IEEEauthorblockN{Arush Gumber}
\IEEEauthorblockA{\textit{IIIT-Delhi} \\
Delhi, India \\
arush23136@iiitd.ac.in}
\and
\IEEEauthorblockN{Anushk Kumar}
\IEEEauthorblockA{\textit{IIIT-Delhi} \\
Delhi, India \\
anushk23115@iiitd.ac.in}

}

\maketitle

\begin{abstract}
The increasing prevalence of lumbar spinal canal stenosis has resulted in a surge of MRI (Magnetic Resonance Imaging), leading to labor-intensive interpretation and significant inter-reader variability, even among expert radiologists. This paper introduces a novel and efficient deep-learning framework that fully automates the grading of lumbar spinal canal stenosis. We demonstrate state-of-the-art performance in grading spinal canal stenosis on a dataset \cite{b1} of 1,975 unique studies, each containing three distinct types of 3D cross-sectional spine images: Axial T2, Sagittal T1, and Sagittal T2/STIR. Employing a distinctive training strategy, our proposed multistage approach effectively integrates sagittal and axial images. This strategy employs a multi-view model with a sequence-based architecture, optimizing feature extraction and cross-view alignment to achieve an AUROC (Area Under the Receiver Operating
Characteristic Curve) of 0.971 in spinal canal stenosis grading surpassing other state-of-the-art methods.
\\The code is available as follows - \url{https://github.com/Deep-learning-exp/M-SCAN/tree/main}

\end{abstract}


\section{Introduction}

Lower back pain is one of the leading causes of long-term disability worldwide. As population age and life expectancy increases, the prevalence of degenerative conditions such as lumbar spinal canal stenosis (SCS) is increasing, becoming a major public health concern. Studies indicate that at least 20\% of the elderly population in India suffers from spinal stenosis, highlighting the urgent need for rapid, effective, and cost-effective diagnostic and monitoring methods. In this condition, the spinal canal narrows to the point at which it can exert pressure on the nerves that travel through the spine. Our research tackles this challenge by developing an advanced deep-learning framework to fully automate the classification of lumbar spinal canal stenosis.

\subsection{Previous Work}
Past work on this uses various methods, including U-Net with multichannel CNN (convolutional neural network) \cite{b2,b3} and 3D/2D CNN \cite{b4}. These approaches \cite{b2,b3,b5} often rely on organized data with uniform histogram images and a uniform number of slices per angle per study, making them overly dependent on specific data types. Our framework is able to handle a variable number of axial and sagittal slices using our multistage approach. We also process the whole spine at once, accounting for histogram and machine-specific differences, providing a holistic prediction for each spine level. We opted for more general data showcasing the real-life difficulties experienced by many radiologists, making it more effective in real-world scenarios.

\begin{figure}[t]
\centerline{\includegraphics[width=70mm,scale=0.5]{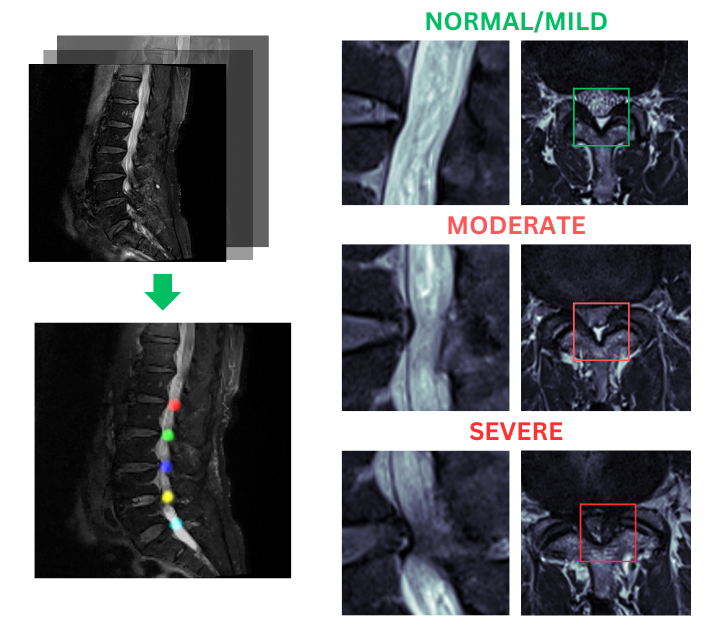}}
\caption{Sample of dataset showing the 3 types of grade present in the dataset, which can be diagnosed using sagittal and axial images.}
\label{fig_data}
\end{figure}

\begin{figure*}[h]
    \centering
    \includegraphics[width=\textwidth,scale=1.0]{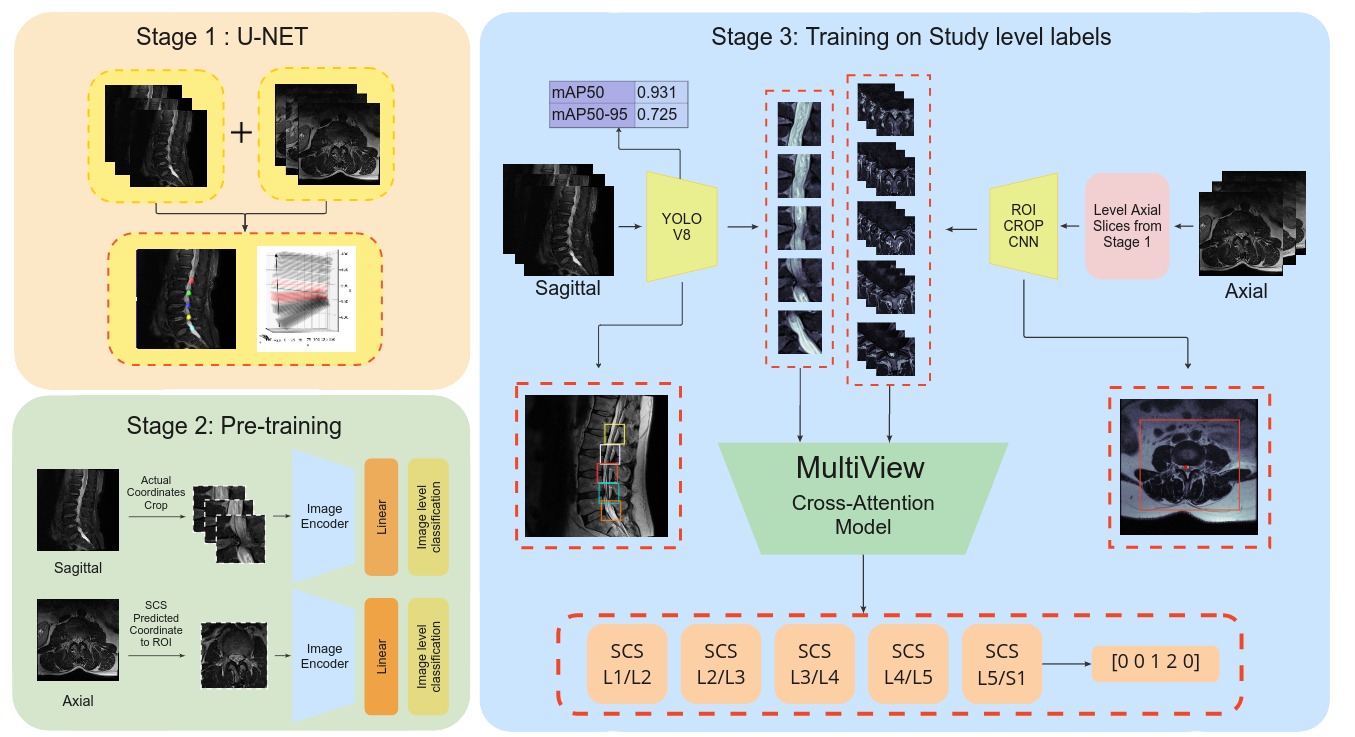}
\caption{The above figure shows our multi-stage framework, giving a brief overview of our process to diagnose spinal canal stenosis using multiple angles of the spine.}
\label{fig_data}
\end{figure*}

\subsection{Our Approach}
For our framework, we used the RSNA 2024 Lumbar Spine Degenerative dataset\cite{b1}, comprising 1,975 studies labeled across five inter-vertebral disc levels (L1/L2, L2/L3, L3/L4, L4/L5, and L5/S1) and categorized into Normal/Mild, Moderate, or Severe SCS. The dataset features Axial T2 images (about 40 slices per study) for horizontal cross-sections, useful for detecting spinal canal narrowing, and Sagittal T2/STIR images (approximately 17 slices per study) for vertical cross-sections, essential for assessing spine alignment, disc degeneration, and nerve root compression. We selected this dataset for its generalizability, as it reflects the variability found in most related datasets. Its diverse image histograms and varying slice counts make it well-suited for real-world applications. Achieving strong metrics here suggests our model's potential to perform well on other MRI datasets. 

Our approach ensures accurate and efficient classification through a three-stage design, offering a comprehensive diagnostic pipeline for analyzing 3D MRI scans.

\begin{itemize}
  \item \textbf{Stage One}: Stage one comprises of training a U-Net \cite{b6} based model to identify SCS points on sagittal images and further developing an algorithm to find the nearby axial slices for each of the five levels using DICOM ( Digital Imaging and Communications in Medicine) metadata.
  \item \textbf{Stage Two}: Based on the precise coordinates provided in the dataset, sagittal and axial images are cropped. A convolutional neural network (CNN) is then trained on these cropped images and image-level labels
  \item \textbf{Stage Three}: In each study, multiple imaging modalities are utilized. Initially, regions of interest (ROIs) are extracted from the YOLO (you only look once) \cite{b7} network and a CNN. A MultiView Cross Attention Model processes these ROIs with five classification heads, each targeting a specific spinal level. 
\end{itemize}

\section{METHODOLOGY}
This section introduces our unique training strategy for diagnosing Spinal Canal Stenosis (SCS), which employs a comprehensive three-stage architecture that includes pre-processing techniques and model architecture design.

\begin{figure*}[h]
    \centering
    \includegraphics[width=\textwidth,scale=1.2]{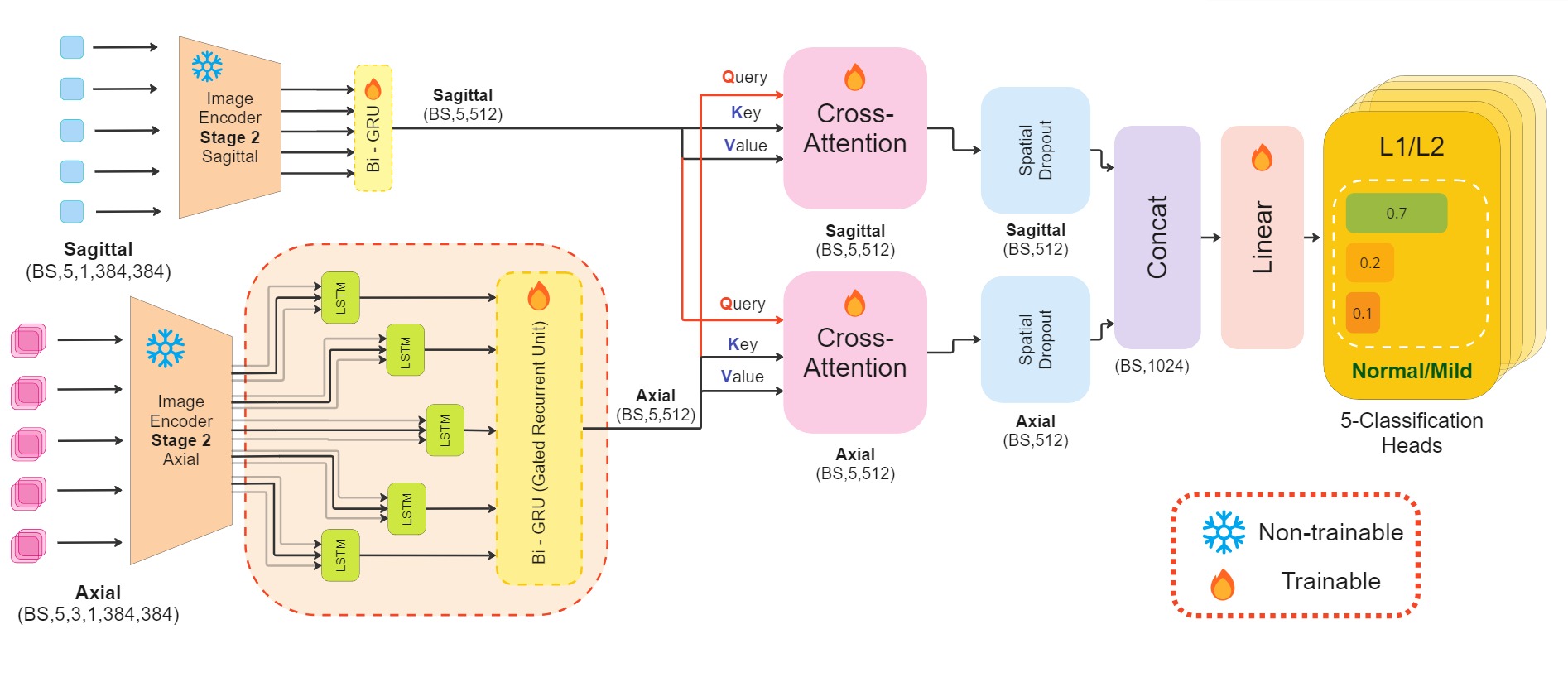}
\caption{The above illustration shows the architecture of our MultiView attention-based network, which includes an image encoder based on the EfficientNet trained in stage two along with our sequence model comprising of LSTM, GRU and attention layers with five classification heads for output.}
\label{fig3}
\end{figure*}

\subsection{Stage One}\label{AA}
Stage one comprises training a U-Net based model, in this stage, we start by extracting 2D coordinates for each spinal level using a U-Net with ResNet50 \cite{b8} as the encoder to identify spinal canal stenosis (SCS) points on sagittal images.
Next, we develop an algorithm to project the 2D coordinates into 3D space, incorporating orientation and position data from the DICOM metadata (ImageOrientationPatient). Then, axial slices are similarly projected into 3D space (Fig. 2). The three nearest axial slices for each of the five spinal levels are identified by calculating the distance between the Z-coordinates of the spinal levels and the axial slices, resulting in a total of 15 axial slices, ensuring comprehensive spine coverage, as shown in stage one of Fig. ~\ref{fig_data}.

\subsection{Stage Two}\label{BB} Stage Two focuses on the pre-training of sagittal and axial slices on image-level labels on cropped data. This process begins by cropping the images based on the precise 2D coordinates of the vertebral levels obtained from the given coordinates for sagittal images and a SCS point prediction model for axial images, as shown in Fig. 2.
For each study, sagittal images are cropped to focus on the five key vertebral levels and the axial images are cropped such that the central canal of the spine is highlighted. Each cropped image is labeled for stenosis severity—Normal/Mild, Moderate, or Severe—according to ground truth data. This classification is performed for each vertebral level to enhance the model’s ability to differentiate stenosis severity. We applied Contrast Limited Adaptive Histogram Equalization (CLAHE) \cite{b9} to enhance localized contrast, facilitating the detection of subtle abnormalities. A CNN based on the EfficientNet \cite{b10} architecture is then trained in a supervised manner using these labeled image crops.

\subsection{Stage Three}
Stage three comprises of using the models and information from the previous two stages and combining them to output a grade for SCS for the five levels of the spine,
\newline

\begin{itemize}
  \item \textbf{Sagittal}: For a 3D cross-section of sagittal images, five crops have to be obtained, each corresponding to the levels -  L1/L2, L2/L3, L3/L4, L4/L5, and L5/S1.

Let \( N \) represent the total number of sagittal slices available for a given subject. The task is to extract information corresponding to the five key spinal levels from these slices. To accomplish this, we employ a YOLOv8 \cite{b11} model that is trained to output probabilities across five distinct classes, each representing one of the spinal levels.

For each sagittal slice \( i \) (where \( i = 1, 2, \ldots, N \)), let \( p_{i,j} \) denote the probability that the \( i \)-th slice corresponds to spinal level \( j \) (where \( j = 1, 2, \ldots, 5 \)). The goal is to identify the optimal sagittal slices corresponding to each spinal level. 
\[
\forall j \in \{1, 2, \ldots, 5\}, \quad \text{select the slice } \, i_j \, \text{such that}
\]
\begin{equation*}
i_j = \arg\max_{i \in \{1, 2, \ldots, N\}} p_{i,j}
\end{equation*}

The selected sagittal slice \( i_j \) will correspond to the slice with the highest probability \( p_{i,j} \) for each spinal level \( j \), ensuring that the most relevant sagittal slice is extracted for each spinal level as shown in Fig.~\ref{fig_data}. 
\newline

\item \textbf{Axial}: For a given patient, the number of axial slices can range from 40 to over 200. To manage this variability, we employed a two-part approach. First, we used the Stage 1 model and algorithm to extract three slices per spinal level, resulting in a total of 15 slices per patient. Subsequently, we applied a Spinal Canal Stenosis (SCS) point prediction model, which utilized a simple CNN encoder with a regression head. This model was designed to crop the slices so that the central canal is prominently highlighted, as illustrated in Fig.~\ref{fig_data}. 
\hfill 
\item \textbf{Model}: The proposed model architecture (illustrated in Fig. 3) is designed for multi-view classification of medical images using sagittal and axial slices. The model employs a dual CNN \cite{b12}architecture for feature extraction, where sagittal and axial images are processed independently through non-trainable stage two pre-trained CNN encoders. These encoders extract 512-dimensional feature representations for each image slice. Formally, 


for a batch size $B$, the sagittal features $\mathbf{h}_{\text{sag}} \in \mathbb{R}^{B \times T \times 512}$ and axial features $\mathbf{h}_{\text{ax}} \in \mathbb{R}^{B \times T \times 3 \times 512}$ are generated where T denotes the sequence length that is 5 here.

After obtaining the image encodings from the CNN, the sagittal embeddings \( \mathbf{E}_s \in \mathbb{R}^{B \times T \times 512} \) are fed directly into a bidirectional Gated Recurrent Unit (Bi-GRU) \cite{b13} layer to capture temporal dependencies within the sagittal cross-section of the spine. The GRU processes these embeddings to produce hidden states \( \mathbf{H}_s \):
\vspace{-1.5em}

\begin{equation*}
\mathbf{H}_s = \text{Bi-GRU}_s(\mathbf{E}_s) \quad \text{where} \quad \mathbf{H}_s \in \mathbb{R}^{B \times T \times 512}
\end{equation*}
\vspace{-1.4em}

For the axial embeddings \( \mathbf{E}_a \in \mathbb{R}^{B \times T \times 3 \times 512} \), where 3 refers to the number of slices per level, these are initially processed by LSTM (long short term memory) \cite{b14} layers. Each LSTM unit handles sequences of three slices per level, which is essential in capturing the sequential nature of axial MRI slices, resulting in hidden states \( \mathbf{H}_{a, i} \):
\vspace{-1.4em}

\begin{equation*}
\mathbf{H}_{a, i} = \text{LSTM}_{a, i}(\mathbf{E}_{a, i}) \quad \text{where} \quad \mathbf{H}_{a, i} \in \mathbb{R}^{B \times T \times 512}
\end{equation*}
\vspace{-1.2em}

The outputs of these LSTM units are then aggregated and passed to a second Bi-GRU layer to capture temporal dependencies across the axial sequences:
\vspace{-0.8em}

\begin{equation*}
\mathbf{E}_{a, \text{bi-gru}} = \text{Bi-GRU}_a(\mathbf{H}_{a}) \quad \text{where} \quad \mathbf{E}_{a, \text{bi-gru}} \in \mathbb{R}^{B \times T \times 512}
\end{equation*}
\vspace{-1.2em}

After processing the embeddings through the GRU and LSTM layers, the attention \cite{b15} mechanism is applied. The output from the axial GRU layer \( \mathbf{O}_{\text{ax}} \in \mathbb{R}^{B \times T \times 512} \) and the output from the sagittal GRU layer \( \mathbf{O}_s \in \mathbb{R}^{B \times T \times 512} \) are used as follows:
\vspace{-0.8em}

\begin{equation*}
\mathbf{O}_1 = \text{Attention}_1(\mathbf{O}_{\text{ax}}, \mathbf{O}_s, \mathbf{O}_s)
\end{equation*}
\vspace{-1.4em}

\noindent
Here, \(\mathbf{O}_1\) is the output of the first cross-attention mechanism where:
\(\mathbf{Q} = \mathbf{O}_{\text{ax}}\) (query), \(\mathbf{K} = \mathbf{O}_s\) (key), and \(\mathbf{V} = \mathbf{O}_s\) (value).


\noindent
Similarly, \(\mathbf{O}_2\) is the output of the second cross-attention mechanism where:
\(\mathbf{Q} = \mathbf{O}_s\), \(\mathbf{K} = \mathbf{O}_{\text{ax}}\), and \(\mathbf{V} = \mathbf{O}_{\text{ax}}\).

\noindent Combining the refined outputs:
\vspace{-0.8em}

\begin{equation*}
\mathbf{O} = \text{concat}(\mathbf{O}_1, \mathbf{O}_2)
\end{equation*}
\vspace{-1.4em}

Following the cross attention \cite{b16} mechanism, the  features \( \mathbf{O}_1 \) and \( \mathbf{O}_2 \) are passed through spatial dropout \cite{b17} and then concatenated to form a comprehensive feature vector \( \mathbf{O} \). This vector is then passed through a fully connected layer to project it into a suitable feature space. The final step involves utilizing separate linear layers and classification heads for each spine level. 
The rationale behind using sequence-based modeling for the full spine instead of direct concatenation is due to patient and machine-specific variations in the histogram, resolution, and number of slices. A sequence-based approach allows for a comprehensive view of the entire spine while accounting for these differences. This method enables predictions to be made at each spine level relative to others, improving accuracy and adaptability across patients.

\end{itemize}

\subsection{Experiments and Results}
The final model was trained on 80\% of the dataset, consisting of approximately 1,580 studies, with the remaining 20\% allocated for the test set. The CNN backbone, a variant of EfficientNet, was pre-trained during the second stage of the pipeline and subsequently frozen during the final training phase. The training was conducted with the AdamW \cite{b18} optimizer, initialized with a learning rate of 1e-4. Given that the lumbar spinal canal stenosis dataset was highly imbalanced, we utilized a weighted cross-entropy (WCE) \cite{b19} loss function during training. The weights were assigned based on the degree of imbalance in the data, ensuring that the model appropriately penalizes misclassifications of minority classes.
\begin{equation*}
L = -\sum_{i=1}^{3} w_i \cdot y_i \cdot \log(\hat{y}_i), \quad \text{where } w = [1, 2, 4]
\end{equation*}
All the models were trained using PyTorch on an NVIDIA 4060 GPU for over 20 epochs. The obtained results are as follows: (AUROC: Area Under the Receiver Operating Characteristic Curve)

\begin{table}[htbp]
\centering
\caption{Comparison of Model Performance}
\begin{tabular}{|l|c|c|c|}
\hline
\textbf{Model} & \textbf{Accuracy} & \textbf{WCE Loss} & \textbf{AUROC} \\
\hline
CNN (Sagittal only) + Concat & 89.82\% & 0.392 & 0.851 \\
\hline

CNN (Sagittal only) + GRU & 91.37\% & 0.336 & 0.863 \\
\hline

CNN (Multi-View) + GRU & 92.60\% & 0.321 & 0.891 \\
\hline

\textbf{M-SCAN (Ours)} & \textbf{93.80\%} & \textbf{0.282} & \textbf{0.971} \\
\hline

\end{tabular}
\vspace{0.5cm}

\begin{tabular}{|l|c|c|}
\hline
\multirow{2}{*}{\textbf{Model}} & \textbf{Binary AUROC} & \textbf{Binary Accuracy} \\
\hline
DeepSpine~\cite{b21} & 0.97 & - \\
\hline

SpineNet~\cite{b20} & 0.95 & - \\
\hline

SpineNetV2~\cite{b3} & - & 93.2\% \\
\hline

Bharadwaj et al.~\cite{b21} & 0.95 & - \\
\hline
Tumko et al.~\cite{b5} & 0.963 & - \\
\hline
\textbf{M-SCAN (Ours)} & \textbf{0.98} & \textbf{95.60\% }\\
\hline
\end{tabular}
\end{table}

\section*{Conclusion}
This study introduces an advanced deep-learning framework that fully automates the classification of lumbar spinal canal stenosis (SCS), achieving remarkable accuracy and efficiency. By employing a multistage approach that integrates both sagittal and axial MRI images through a sequence-based architecture and multi-view cross-attention mechanism, our model has demonstrated state-of-the-art performance in predicting the severity of SCS across multiple spinal levels. The model’s ability to achieve a predictive precision of 93.80\% and an AUROC of 0.971 surpasses existing benchmarks, highlighting the efficacy of our approach.

The robustness of the framework is demonstrated not only through its high predictive accuracy but also by its adaptability to variations in image resolution, histogram distribution, and slice counts across different patients. This adaptability is a direct result of our sequence-based modeling strategy, which enables the model to consider the entire spine comprehensively while accounting for patient-specific and machine-specific variances. 
By providing a reliable, scalable, and fully automated solution, this work has the potential to greatly enhance the diagnostic process for spinal stenosis, ultimately improving patient outcomes in clinical practice. Future research will focus on further refining this model, exploring its applicability to other spinal conditions, and integrating it into clinical workflows for real-time diagnosis and decision support.


@ARTICLE{dataset,
    author = {Tyler Richards, Jason Talbott, Robyn Ball, Errol Colak, Adam Flanders, Felipe Kitamura, John Mongan, Luciano Prevedello, Maryam Vazirabad.},
    title = {RSNA 2024 Lumbar Spine Degenerative Classification},
    publisher = {Kaggle},
    year = {2024},
    url = {https://kaggle.com/competitions/rsna-2024-lumbar-spine-degenerative-classification}
}

@ARTICLE{lu2018deepspineautomatedlumbarvertebral,
      title={DeepSPINE: Automated Lumbar Vertebral Segmentation, Disc-level Designation, and Spinal Stenosis Grading Using Deep Learning}, 
      author={Jen-Tang Lu and Stefano Pedemonte and Bernardo Bizzo and Sean Doyle and Katherine P. Andriole and Mark H. Michalski and R. Gilberto Gonzalez and Stuart R. Pomerantz},
      year={2018},
      eprint={1807.10215},
      archivePrefix={arXiv},
      primaryClass={cs.CV},
      url={https://arxiv.org/abs/1807.10215}, 
}



\begin{thebibliography}{00}

\bibitem{b1}T. Richards, J. Talbott, R. Ball, E. Colak, A. Flanders, F. Kitamura, J. Mongan, L. Prevedello, and M. Vazirabad, ``RSNA 2024 Lumbar Spine Degenerative Classification,'' Kaggle, 2024. [Online]. Available: https://kaggle.com/competitions/rsna-2024-lumbar-spine-degenerative-classification.

\bibitem{b2}Jen-Tang Lu, Stefano Pedemonte, Bernardo Bizzo, Sean Doyle, Katherine P. Andriole, Mark H. Michalski, R. Gilberto Gonzalez, Stuart R. Pomerantz, ``DeepSPINE: Automated Lumbar Vertebral Segmentation, Disc-level Designation, and Spinal Stenosis Grading Using Deep Learning,'' Machine Learning for Healthcare (MLHC) 2018.

\bibitem{b3}Windsor, R., et al. ''SpineNetV2: Automated Detection, Labelling and Radiological Grading of Clinical MR Scans''. ArXiv, 2022.

\bibitem{b4}Kim, J.K., Chang, M.C. ''Convolutional Neural Network Algorithm Trained on Lumbar Spine Radiographs to Predict Outcomes of Transforaminal Epidural Steroid Injection for Lumbosacral Radicular Pain from Spinal Stenosis''. Sci Rep 14, 8490 (2024). 

\bibitem{b5}Tumko, V., Kim, J., Uspenskaia, N. et al. "A neural network model for detection and classification of lumbar spinal stenosis on MRI". Eur Spine J 33, 941–948 (2024). https://doi.org/10.1007/s00586-023-08089-2


\bibitem{b6}Olaf Ronneberger, Philipp Fischer, Thomas Brox. ''U-Net: Convolutional Networks for Biomedical Image Segmentation,'' MICCAI 2015.

\bibitem{b7}J. Redmon, S. Divvala, R. Girshick and A. Farhadi, ``You Only Look Once: Unified, Real-Time Object Detection,'' 2016 IEEE Conference on Computer Vision and Pattern Recognition (CVPR), Las Vegas, NV, USA, 2016, pp. 779-788, doi: 10.1109/CVPR.2016.91.

\bibitem{b8}K. He, X. Zhang, S. Ren and J. Sun, ``Deep Residual Learning for Image Recognition,'' 2016 IEEE Conference on Computer Vision and Pattern Recognition (CVPR), Las Vegas, NV, USA, 2016, pp. 770-778, doi: 10.1109/CVPR.2016.90.

\bibitem{b9}S. M. Pizer, R. E. Johnston, J. P. Ericksen, B. C. Yankaskas and K. E. Muller, ``Contrast-Limited Adaptive Histogram Equalization: Speed and Effectiveness,'' [1990] Proceedings of the First Conference on Visualization in Biomedical Computing, Atlanta, GA, USA, 1990, pp. 337-345, doi: 10.1109/VBC.1990.109340.

\bibitem{b10}Mingxing Tan, Quoc V. Le, ``EfficientNet: Rethinking Model Scaling for Convolutional Neural Networks,'' ICML 2019.

\bibitem{b11}Jocher, G., Chaurasia, A.,  Qiu, J. (2023). ''Ultralytics YOLO (Version 8.0.0) [Computer Software],'' https://github.com/ultralytics/ultralytics.

\bibitem{b12}S. Albawi, T. A. Mohammed and S. Al-Zawi, ``Understanding of a Convolutional Neural Network,'' 2017 International Conference on Engineering and Technology (ICET), Antalya, Turkey, 2017, pp. 1-6, doi:

\bibitem{b13}Kyunghyun Cho, Bart van Merriënboer, Caglar Gulcehre, Dzmitry Bahdanau, Fethi Bougares, Holger Schwenk, and Yoshua Bengio. 2014. ''Learning Phrase Representations Using RNN Encoder–Decoder for Statistical Machine Translation,'' In Proceedings of the 2014 Conference on Empirical Methods in Natural Language Processing (EMNLP), pages 1724–1734, Doha, Qatar. Association for Computational Linguistics.

\bibitem{b14}Sepp Hochreiter and Jürgen Schmidhuber. 1997. ''Long Short-Term Memory,'' Neural Comput. 9, 8 (November 15, 1997), 1735–1780.

\bibitem{b15}Ashish Vaswani, Noam Shazeer, Niki Parmar, Jakob Uszkoreit, Llion Jones, Aidan N. Gomez, Łukasz Kaiser, and Illia Polosukhin. 2017. ''Attention is All You Need,'' In Proceedings of the 31st International Conference on Neural Information Processing Systems (NIPS'17).

\bibitem{b16}Mozhdeh Gheini, Xiang Ren, and Jonathan May. 2021. ''Cross-Attention is All You Need: Adapting Pretrained Transformers for Machine Translation,'' In Proceedings of the 2021 Conference on Empirical Methods in Natural Language Processing, pages 1754–1765, Online and Punta Cana, Dominican Republic. Association for Computational Linguistics.

\bibitem{b17}Tompson, J., Goroshin, R., Jain, A., LeCun, Y., Bregler, C. (2014). ''Efficient Object Localization Using Convolutional Networks,'' 2015 IEEE Conference on Computer Vision and Pattern Recognition (CVPR), 648-656.


\bibitem{b18}Loshchilov, I., Hutter, F. (2017). ''Decoupled Weight Decay Regularization,'' International Conference on Learning Representations.

\bibitem{b19}Ö. Özdemir and E. B. Sönmez, ``Weighted Cross-Entropy for Unbalanced Data with Application on COVID X-ray Images,'' 2020 Innovations in Intelligent Systems and Applications Conference (ASYU), Istanbul, Turkey, 2020, pp. 1-6, doi: 10.1109/ASYU50717.2020.9259848.

\bibitem{b20}Amir Jamaludin, Timor Kadir, Andrew Zisserman. ''SpineNet: Automated Classification and Evidence Visualization in Spinal MRI,'' (2017). Medical Image Analysis 41, p 63-73.

\bibitem{b21}Upasana Upadhyay Bharadwaj, Miranda Christine, Steven Li, Dean Chou, Valentina Pedoia, Thomas M. Link, Cynthia T. Chin, Sharmila Majumdar. ''Deep Learning for Automated, Interpretable Classification of Lumbar Spinal Stenosis and Facet Arthropathy from Axial MRI,'' Imaging Informatics and Artificial Intelligence 2023.

\bibitem{b22}Qin, Jiuming Liu, Che Cheng, Sibo Guo, Yike Arcucci, R. (2024). ''Freeze the Backbones: a Parameter-Efficient Contrastive Approach to Robust Medical Vision-Language Pre-Training,'' 1686-1690. 10.1109/ICASSP48485.2024.10447326.

\bibitem{b23}Rizwan Qureshi, Mohammed Gamal Ragab, Said Jadid Abdulkader, et al. ''A Comprehensive Systematic Review of YOLO for Medical Object Detection (2018 to 2023),'' TechRxiv, July 17, 2023.

\bibitem{b24}Arman Avesta, Sajid Hossain, MingDe Lin, Mariam Aboian, Harlan M. Krumholz, and Sanjay Aneja, ``Comparing 3D, 2.5D, and 2D Approaches to Brain Image Segmentation,'' medRxiv, Cold Spring Harbor Laboratory Press, 2022, doi: 10.1101/2022.11.03.22281923.






\end{thebibliography}
\end{document}